\documentclass[aps,prl,reprint,superscriptaddress,nofootinbib,floatfix,longbibliography]{revtex4-2}

\usepackage{amsmath}
\usepackage{bm}
\usepackage{amsfonts}
\usepackage{graphicx}
\usepackage{hyperref}
\usepackage{soul}
\usepackage{color}

\enlargethispage{\baselineskip}

\hypersetup{
     colorlinks   = true,
     citecolor    = blue
}
\everymath{\displaystyle}

\newcommand{\gBL}{g_{B\textrm{--}L}}
\newcommand{\mA}{m_{\scriptscriptstyle A'}}

\newcommand{\apjl}{Astrophys.~J.~Lett.}

\begin{document}

\title{Neutrino astronomy as a probe of physics beyond the Standard Model:\\ decay of sub-MeV $B$-$L$ gauge boson dark matter}

\newcommand{\TDLI}{\affiliation{Tsung-Dao Lee Institute (TDLI) \& School of Physics and Astronomy, Shanghai Jiao Tong University, \\ Shengrong Road 520, 201210 Shanghai, P.\ R.\ China}}

\author{Weikang Lin}
\email{weikanglin@sjtu.edu.cn}
\TDLI

\author{Luca Visinelli}
\email{luca.visinelli@sjtu.edu.cn}
\TDLI

\author{Donglian Xu}
\email{donglianxu@sjtu.edu.cn}
\TDLI

\author{Tsutomu T. Yanagida}
\email{tsutomu.tyanagida@sjtu.edu.cn \newline}
\TDLI
\affiliation{Kavli IPMU (WPI), The University of Tokyo, Kashiwa, Chiba 277-8583, Japan}
\date{\today}

\begin{abstract}
The $U(1)_{B\textrm{--}L}$ symmetry, the essential component in the seesaw mechanism and leptogenesis, is naturally equipped with a massive gauge boson. If this gauge boson is the dark matter, the scenario consistent with the seesaw mechanism predicts the gauge coupling to be of the order of $\mathcal{O}(10^{-19})$ for masses $\lesssim1$\,MeV, dominantly decaying into active neutrinos. We stress and explore the important role of astrophysical neutrinos of energies from $\mathcal{O}(1)$\,keV to $\sim1$\,MeV in testing the well-motivated $B$-$L$ symmetry extension to the Standard Model, which has been missed in the literature to date. Compared to other dark matter models, the neutrino flux in the sub-MeV energy range is a unique prediction in our setup and, once detected, would serve as a smoking gun for the existence of this $B$-$L$ gauge boson and its role as the dark matter particle, opening new windows to tackle cosmological and astrophysical conundra.
\end{abstract}

\maketitle

{\bf Introduction\;---} Dark matter (DM) could be in the form of a light boson field~\cite{Preskill:1982cy,Abbott:1982af,Dine:1982ah} such as a Nambu-Goldstone boson~\cite{Nambu:1960tm,Goldstone:1961eq} or a massive dark photon~\cite{Okun:1982xi,Georgi:1983sy,Fabbrichesi:2020wbt}. The most famous example for the former case is the QCD axion~\cite{Weinberg:1977ma, Wilczek:1977pj}. On the other hand, a new, massive dark photon could arise in models that possess an Abelian $B$-$L$ gauge symmetry~\cite{Georgi:1973}, one of the most natural and well motivated among the proposed Abelian gauge symmetries.

The $U(1)_{B\textrm{--}L}$ symmetry naturally requires three right-handed neutrinos to cancel out gauge anomalies; its spontaneous breaking generates large Majorana masses for the right-handed neutrinos, $N_i$, where $i\in\{1,2,3\}$ denotes the neutrino generation. The presence of heavy Majorana neutrinos is the key point for inducing small masses of the active neutrinos through the seesaw mechanism~\cite{Yanagida:1979as, *Yanagida:1979gs,GellMann:1980vs,Minkowski:1977sc,Wilczeck:1979CP} and generating the baryon asymmetry in the universe via the leptogenesis mechanism~\cite{Fukugita:1986hr, Buchmuller:2005eh}.

The $B$-$L$ breaking scale $V$ as inferred from the observed small neutrino masses and the scale of the leptogenesis is of the order of $V=(10^{12}\,\textrm{--}\,10^{16})\,$GeV. On the other hand, the mass of the $B$-$L$ gauge boson, $A'$, depends on the gauge coupling constant of the $U(1)_{B\textrm{--}L}$ theory, $\gBL$, as $\mA =2\gBL V$. The estimated range of $V$ comes from the seesaw mechanism, for which the neutrino mass is given by $m_\nu \simeq \tfrac{g^2}{G}\tfrac{\langle H\rangle^2}{V}$, where $\langle H\rangle\sim100\,$GeV is the vacuum expectation value of the Higgs field and $g, G$ are Yukawa coupling constants. Matching the expression with the mass of the heaviest neutrino $\sim \mathcal{O}(10^{-2})\,$eV gives $V\sim10^{15}\,$GeV for $G\sim1$ and $g\sim1$, while a smaller $g\sim 1/30$ would lead to $V\sim10^{12}\,$GeV. Considering the uncertainty of $G$ appearing in the relation between $V$ and the right-handed neutrino mass $(=GV)$, we take $V=(10^{12}\,\textrm{--}\,10^{16})\,$GeV.

Motivated by the excess reported by the XENON1T collaboration~\cite{XENON:2020rca}, one of us recently pointed out that the dark photon $A'$ can be the DM particle if its mass lies within the range $\mA=\mathcal{O}(10)\,$keV~\cite{Choi:2020kch, Okada:2020evk}. This dark photon decays mainly to two neutrinos, since the decay to three photons is extremely suppressed in the region considered here~\cite{Choi:2020kch,Okada:2020evk}. Here, we name this gauge dark photon DM the ``$B$-$L$ f{\'e}eton'' or f{\'e}eton for short.\footnote{The small coupling constant $\gBL$ gives both a sub-MeV gauge boson mass and a small decay width even for a large symmetry breaking scale $V$. Here, we name this the ``f{\'e}eton mechanism.'' The name ``f{\'e}eton'' comes from the French word {\it f{\'e}e} for fairy.} Mechanisms leading to the production in the early universe of $B$-$L$ f{\'e}eton as the dominant DM component are discussed in Ref.~\cite{Choi:2020dec}.\footnote{Alternatively, the f{\'e}eton DM can be produced from inflationary fluctuations through the mechanism discussed in Ref.~\cite{Graham:2015rva}. In this scenario, the f{\'e}eton mass is related to the Hubble scale of inflation $H_I$ by $m_{A'}\approx1\,{\rm keV}\times(10^{12}\,{\rm GeV}/H_I)^4$. When $m_A'=\mathcal{O}(10)\,$keV, we have $H_I\lesssim10^{12}\,{\rm GeV}<V$ which is consistent with the $B$-$L$ symmetry being broken before or during inflation.} Addressing the XENON1T anomaly requires a large kinetic mixing ($\kappa\sim10^{-15}$) between the $B$-$L$ gauge boson and the photon~\cite{Choi:2020kch, Okada:2020evk}, which is difficult to be generated by one-loop diagrams of electron with the gauge coupling assumed ($\gBL\sim10^{-16}-10^{-18}$).

In this work, we point out that a more natural and the most important prediction of this model is the predominant decay channel into neutrinos with an energy of $\mathcal{O}(1)$\,keV to $\sim1$\,MeV. Although previous work considered the phenomenology of a DM particle decaying predominantly into a neutrino pair~\cite{Palomares-Ruiz:2007egs,McKeen:2018xyz,Bondarenko:2020vta,Garcia-Cely:2017oco, Chacko:2018uke}, a potentially detectable neutrino flux at these energies is a unique feature in our model to date. For the first time, we show the important role of astrophysical neutrinos within such an energy range in testing the well-motivated $B$-$L$ symmetry extension to the Standard Model of particle physics (SM), which is otherwise difficult to probe with high-energy particle colliders.
%
%
We investigate the spectrum and the morphology (i.e., the directional dependence) of low-energy neutrino signals from f{\'e}eton DM decays, showing that neutrino astronomy could provide a promising window to probe the particle content beyond SM, a novel and independent test on cosmological models, and a way to study the DM profile of our Milky Way (MW). Throughout the paper we adopt natural units with $c=\hbar=1$.

{\bf Neutrino signals from f{\'e}eton DM decays\;---}
Here, we calculate the neutrino flux from f{\'e}eton decay, assuming it constitutes the totality of DM. In our model, the f{\'e}eton mainly decays into left-handed neutrinos via the interaction $\mathcal{L}_{\rm int}=\gBL\bar{\nu}_{\rm L}\gamma^\mu \nu_{\rm L} A'_{\mu}$ with a decay width given by (see e.g.\ Refs.~\cite{Ilten:2018crw,Fabbrichesi:2020wbt})
\footnote{ Other models, such as majoron DM, only predict a small neutrino flux within the $\mathcal{O}(1)$\,keV to $\mathcal{O}(100)$\,keV energy window. According to Eq.~(13) in Ref.~\cite{Garcia-Cely:2017oco}, the strongest neutrino flux from the decay of majoron DM of mass $100\,$keV is at least about $2$ orders of magnitude smaller than our most optimistic prediction. Moreover, majoron models are plagued by the arguments that quantum gravity violates global symmetry~\cite{Banks:2010zn,Witten:2017hdv,Harlow:2018tng,Alvey:2020nyh}.}
\begin{equation}
    \label{eq:Gamma}
    \Gamma_{A'} = \frac{1}{8\pi}\gBL^2 \mA\,.
\end{equation}
Cosmological observations, such as the Integrated Sachs Wolfe effect on the cosmic microwave background, bound the DM lifetime to be $\tau_{A'} \equiv 1/\Gamma_{A'} \gtrsim 150\,$Gyr, when assuming a single decaying DM component~\cite{DeLopeAmigo:2009dc, Audren:2014bca, DES:2020mpv, Enqvist:2019tsa}. On the other hand, megaparsec- and galaxy-scale structures constrain the mass of thermally-produced (warm) DM particles (WDM), see e.g.\ Refs.~\cite{Palanque-Delabrouille:2019iyz, Gilman:2019nap, Garzilli:2021qos}. Assuming f{\'e}eton DM was thermally produced, we conservatively adopt $\mA\gtrsim1.9\,$keV~\cite{Garzilli:2021qos}, although the constraint can be somewhat stronger in other studies. Note, that this WDM mass constraint can differ in models in which f{\'e}eton DM is produced non-thermally~\cite{Choi:2020kch}.

\begin{figure}[tbp]
    \centering
    \includegraphics[width=0.99\linewidth]{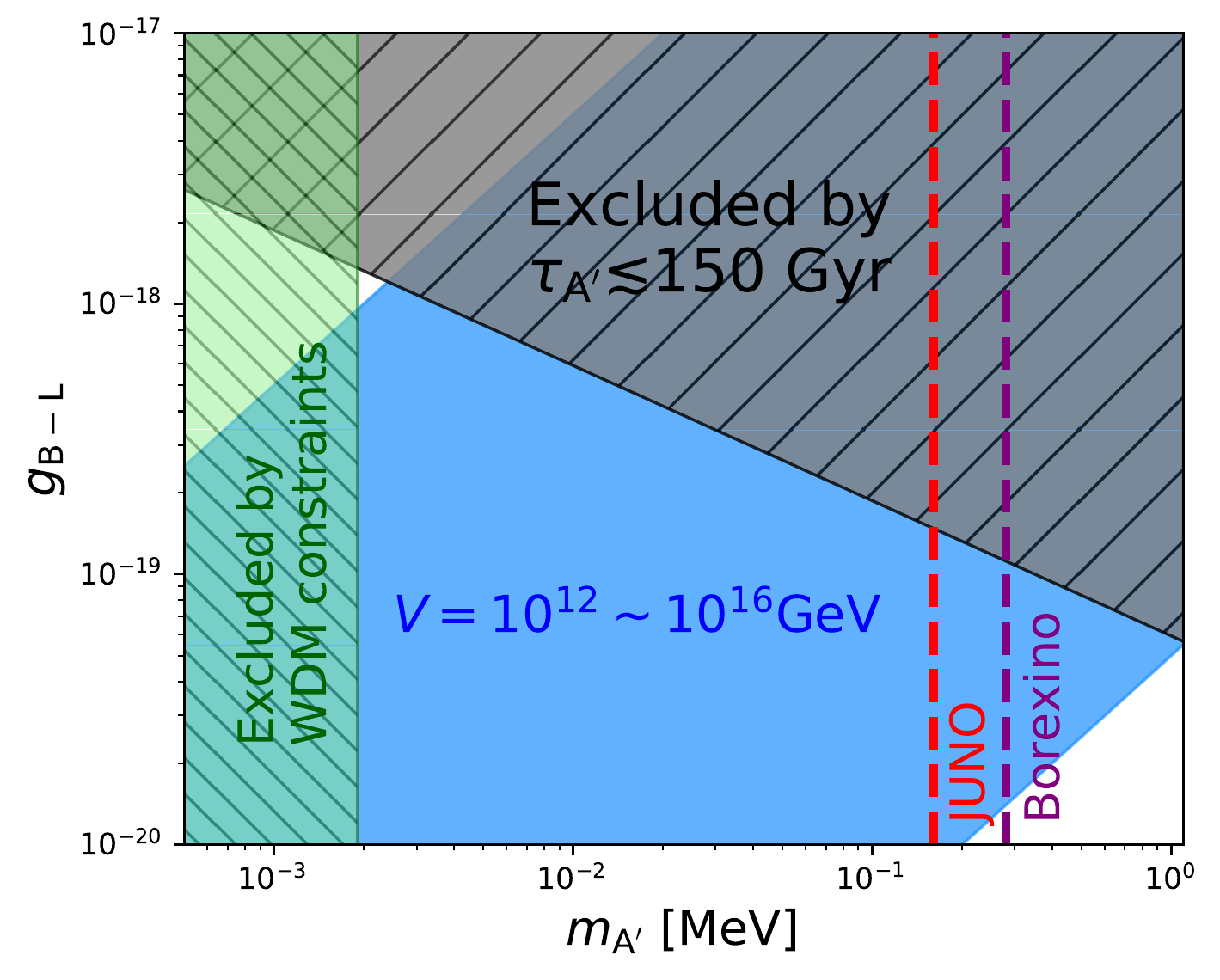}
    \caption{Model parameter space $(\mA, \gBL)$. The light blue band shows the region for which $V=(10^{12}\,\textrm{--}\,10^{16})\,$GeV, as suggested from the smallness of the neutrino masses and the scale of leptogenesis. Cosmological considerations exclude the gray region ($1/\Gamma_{A'}\lesssim150\,$Gyr) and the green region ($\mA\lesssim1.9\,$keV).  The dashed lines show the thresholds of the current Borexino (purple) and the future JUNO (red) experiments. For JUNO, we estimate a future detection of up to $\sim2300$ events per year caused by neutrinos from f{\'e}eton decays, see the main text for details.}
    \label{fig:ParamSpace}
\end{figure}

Fig.~\ref{fig:ParamSpace} shows the region of the parameter space that is motivated by theory (light blue band), along with the  cosmological and astrophysical constraints presented above. Remarkably, cosmological constraints alone yield $V\gtrsim10^{12}\,$GeV, consistently with the suggestions from the scale of the heaviest neutrino mass and leptogenesis. Moreover, the viable mass range $1.9\,{\rm keV}\lesssim\mA\lesssim1$\,MeV is consistent with the light mass condition required for electron decay suppression and with the constraints for the mass of a WDM particle~\cite{Choi:2020kch,Choi:2020dec}. Also shown is the thresholds of the Borexino detector (vertical purple dashed line)~\cite{BOREXINO:2014pcl} and the Jiangmen Underground Neutrino Observatory (JUNO) (vertical red dashed line)~\cite{JUNO:2015zny}.

The search for by-product particles from DM annihilation or decay dates back to more than 40 years ago~\cite{Silk:1984zy,1990Natur.346...39L}. Here, we assess the neutrino signals by separately calculating the flux from f{\'e}eton decay within the Galactic DM halo and from the isotropic extragalactic background. For the Galactic (Gal) signal, the differential (number) flux for \emph{each} neutrino species from f{\'e}eton decay is (see e.g.\ Refs.~\cite{Asaka:1998ju,Bergstrom:1997fj,Blasi:2001hr, Cirelli:2010xx})
\begin{equation}
	\label{eq:galflux}
	\frac{{\rm d}^2\Phi_{\nu}^{\rm {\textsc{g}al}}}{{\rm d}E_\nu\,{\rm d}\Omega} = \frac{1}{3} \frac{\Gamma_{A'}}{4\pi \mA}\,\frac{{\rm d}N_\nu}{{\rm d}E_\nu}\,D(b,\ell)\,,
\end{equation}
where the (nearly) monochromatic neutrinos emerge with the energy spectrum ${\rm d}N_\nu/{\rm d}E_\nu \simeq 2\delta(E_\nu - \mA/2)$, while the astrophysical D-factor computed along the line of sight (l.o.s.) $s$ with galactic coordinates $(b, \ell)$ is
\begin{equation}
    \label{eq:Gal-nu-flux}
	D(b,\ell)=\int_{\rm l.o.s.} {\rm d}s\,\rho_{\textsc{a}'}(r(b, \ell))\,.\\
\end{equation}
Here, $r(b, \ell) = (s^2+r_\odot^2-2r_\odot s \cos b\cos\ell)^{1/2}$ is the distance with respect to the Galactic center (GC) and $r_\odot$ is the distance between the Sun and GC. We consider two distinct DM density distributions $\rho_{\textsc{a}'}(r)$ in MW: {\it i)} a Navarro–Frenk–White (NFW) profile~\cite{Navarro:1995iw} of density scale $\rho_s = 1.4\times 10^7 M_\odot{\rm\,kpc^{-3}}$ and scale radius $r_s = 16\,$kpc; {\it ii)} a cored Burkert profile~\cite{Burkert:1997fz} of density scale $\rho_s = 4.1\times 10^7 M_\odot{\rm\,kpc^{-3}}$ and scale radius $r_s = 9\,$kpc~\cite{Nesti:2013uwa}. Encoding this modeling and Eq.~\eqref{eq:Gamma} leads to the Galactic (angular) differential flux with a monochromatic energy at $E_\nu=\mA/2$, which reads
\begin{equation}
    \frac{{\rm d}\Phi_{\nu}^{\rm {\textsc{g}al}}}{{\rm d}\Omega} = \frac{\gBL^2\,\rho_s\,r_s}{48\pi^2}\tilde{D}(\cos\theta)\,,
\end{equation}
where $\cos\theta = \cos b\cos\ell$ is the angle from GC and $\tilde{D}(\cos\theta) \equiv D(b,\ell)/(\rho_s r_s)$ represents the morphology of the Galactic signal and depends on the DM halo profile. For an NFW profile, we have
\begin{equation}\label{eq:galfulx_result}
    \frac{{\rm d}\Phi_{\nu}^{\rm {\textsc{g}al}}}{{\rm d}\Omega}= 7.9\times10^5\tilde{D}_{\rm N}(\cos\theta)\left(\frac{\gBL}{10^{-19}}\right)^2\![{\rm cm}^{-2}\,{\rm s}^{-1}\,{\rm sr}^{-1}],
\end{equation}
where $\tilde{D}_{\rm N}\propto\tilde{D}$ is normalized so that $\smallint \tilde{D}_{\rm N}\,{\rm d}\Omega=4\pi$. It is worth pointing out that the total flux is independent of $\mA$, which will allow us to directly infer the gauge coupling constant from the flux amplitude. Using a Burkert profile would lead to a different morphology function $\tilde{D}_{\rm N}$ (see the description of Fig.~\ref{fig:NuFlux} below), as well as a factor of $\sim0.92$ difference (smaller) in the total flux $\Phi_\nu$.
\begin{figure}[tbp]
    \centering
    \includegraphics[width=\linewidth]{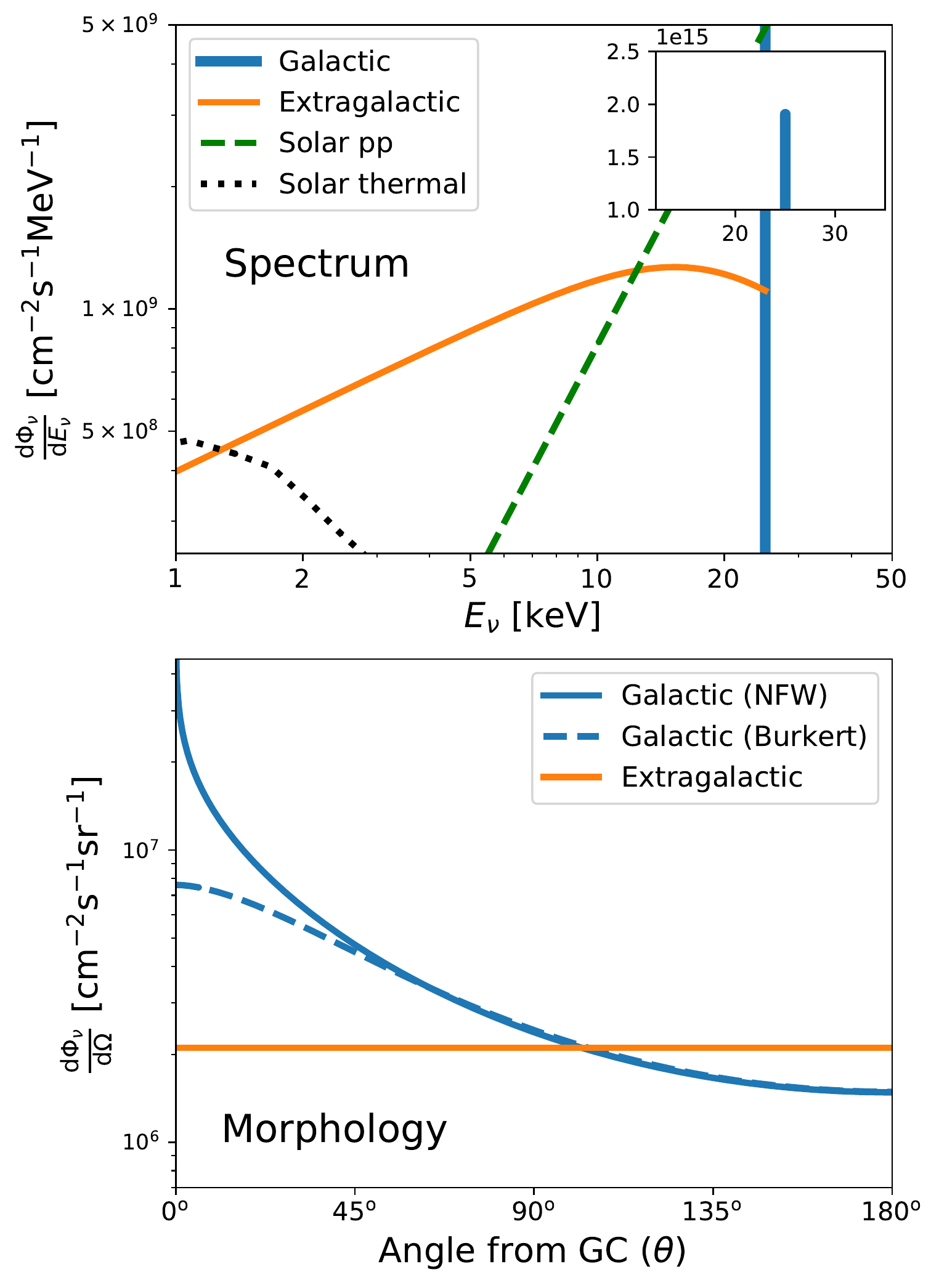}
    \caption{The neutrino fluxes from the decay of f{\'e}eton DM. Top: the full-sky energy spectra of the fluxes for the Galactic (blue) and the extragalactic (orange) signals. Note, the total Galactic flux actually is about $1.5$ times larger than the extragalactic one, although the peak is high due to its small energy dispersion induced by the DM velocity dispersion.  For a comparison, we plot some dominant solar (electron) neutrino backgrounds in green-dashed (pp) and black-dotted (thermal source processes) curves. Bottom: the morphology of the fluxes integrated over the energy. The solid (dashed) blue curve is the Galactic signal assuming an NFW (Burkert) DM profile. The extragalactic signal shown in orange is isotropic. Here, $\mA=50\,$keV and $\gBL=2\times10^{-19}$.}
    \label{fig:NuFlux}
\end{figure}

For the extragalactic (eg) signal, we consider a homogeneous DM distribution. Taking into account the time dilation and redshift of the neutrino energy, the extragalactic differential flux becomes~\cite{Kawasaki:1997ah}
\begin{align}
	\label{eq:egflux}
    \frac{{\rm d}^2\Phi_{\nu}^{\rm eg}}{{\rm d}E_\nu\,{\rm d}\Omega}& =\frac{1}{3}\int_0^{+\infty}\!\!{\rm d}z\,\frac{\Gamma_{A'}\,\rho_{\textsc{a}'}^0}{4\pi H(z)\,\mA}\frac{{\rm d}N}{{\rm d}E'}\bigg|_{E_\nu'=E_\nu(1+z)}\\
    &=\!\frac{\gBL^2\Omega_{\rm A'}H_0}{64\pi^3\mA\ell_{\rm Pl}^2}\mathcal{F}\Big(\frac{2E_\nu}{\mA}\Big)\,,\nonumber\\
    \mathcal{F}(x)&=\Big(\frac{x}{1-\Omega_{\Lambda}+\Omega_{\Lambda}x^3}\Big)^{1/2}\,,
\end{align}
where $H(z)$ is the Hubble rate at redshift $z$, $\ell_{\rm Pl} \approx 1.62\times 10^{-33}\,$cm is the Planck length, $\rho_{\textsc{a}'}^0$ is the present cosmological DM density, and in the last step we have used Eq.~\eqref{eq:Gamma} and we introduced the cosmological fraction from $\rho_{A'}=3\Omega_{A'}H_0^2/(8\pi\ell_{\rm Pl}^2)$.\footnote{Since the integral is dominated by redshifts $z\lesssim 10$, we ignore the neutrino optical depth and contribution to the evolution of $H(z)$ from radiation and the change in the DM content due to decay. We have also ignored the DM proper motion over cosmological distances.} Identifying $\Omega_{\rm A'}h=\Omega_{\rm DM}h \approx 0.176$ from the {\it Planck} 2018 release~\cite{Planck:2018vyg}, we obtain the extragalactic (energy and angular) differential flux
\begin{eqnarray}
    \label{eq:flux-prefactor}
    \frac{{\rm d}^2\Phi_{\nu}^{\rm eg}}{{\rm d}E_\nu\,{\rm d}\Omega} &=& 1.1\times10^7\,\left(\frac{\gBL}{10^{-19}}\right)^2\Big(\frac{0.1{\rm\,MeV}}{\mA}\Big)\\
    && \times\mathcal{F}\Big(\frac{2E_\nu}{\mA}\Big)\,[{\rm cm}^{-2}\,{\rm s}^{-1}\,{\rm sr}^{-1}\,{\rm MeV}^{-1}]\,.\nonumber
\end{eqnarray}
In the expressions above, the function $\mathcal{F}$ depends on the cosmological model; here, we assume the standard cosmological model to take place. Setting $\Omega_\Lambda=0.69$ and integrating over the neutrino energy yields
\begin{equation}\label{eq:eg_flux_total}
    \frac{{\rm d}\Phi_{\nu}^{\rm eg}}{{\rm d}\Omega} = 5.2\times10^5\,\left(\frac{\gBL}{10^{-19}}\right)^2\,[{\rm cm^{-2}\,s^{-1}\,sr^{-1}}]\,.
\end{equation}

The spectra and the morphology of both Galactic and extragalactic neutrino signals are shown in Fig.~\ref{fig:NuFlux}. For the monochromatic Galactic signal, we smear the spectrum with an energy dispersion induced by the DM velocity dispersion ($\sigma_v\sim100\,$km/s) in the Galactic halo. Note, that the Galactic integrated flux $(\Phi_\nu^{\rm \textsc{g}al})$ is only about $1.5$ times larger than the extragalactic one $(\Phi_\nu^{\rm eg})$. Interestingly, the extragalactic signal also has a peak in the energy spectrum. We shall later discuss the cosmological application of such a feature. The bottom panel of Fig.~\ref{fig:NuFlux} shows the distribution of the differential flux with respect to the angle $\theta$. Clearly, the extragalactic component is isotropic (orange solid line), while the Galactic component peaks near the direction of GC. Near the GC region, the shape of the Galactic signal depends on the DM halo profile used, with the NFW profile predicting a spiked distribution of the neutrino signal (blue solid line), while the Burkert profile producing a cored distribution (blue dashed line).

Importantly, with the theoretically and observationally consistent parameter space, the f{\'e}eton DM scenario predicts a non-negligible neutrino signal. For a comparison, we show in the upper panel of Fig.\,\ref{fig:NuFlux} the solar pp neutrino flux with a modeling uncertainty of $\sim$ 1\%~\cite{Bahcall:1997eg,Bahcall:2004mz,Haxton:2012wfz,OHare:2015utx} and those produced in thermal source processes with an uncertainty of $\sim 10\%$~\cite{Vitagliano:2017odj}. At the corresponding energy range, the neutrino flux from the extragalactic f{\'e}eton decay can be comparable to the solar backgrounds, although the solar pp neutrinos dominate at higher energies.
The smooth angular dependence of the neutrino signal from f\'eeton decay makes it more discernible from other localized astrophysical sources once directional detections are implemented.

The discussions above motivate the future developments of low-energy-threshold neutrino detectors targeting astrophysical neutrinos between $\mathcal{O}(1)$\,keV and $\mathcal{O}(100)$\,keV. The proposed goal may be within reach in the coming years. Indeed, the detection threshold of the recoil electron in the Borexino experiment  is $E_{\rm th}=50$\,keV~\cite{BOREXINO:2014pcl}, which corresponds to a minimum neutrino energy of $\sim140$\,keV or a f{\'e}eton mass of $\sim280$\,keV. Within JUNO, consisting of a $20$ kiloton multi-purpose underground liquid scintillator detector~\cite{JUNO:2015zny} with a potential $20$\,keV detection threshold~\cite{JUNO_MM_trigger}, it is promising to survey the f{\'e}eton parameter space in the high mass end. We calculate the electron recoil spectrum and estimate a future JUNO detection of up to $2300$ events per year caused by neutrinos from f{\'e}eton decays, assuming a f\'eeton mass $m_A'=300$\,keV.\footnote{Note, that the f{\'e}eton decays into all neutrino and anti-neutrino flavors.}

At sub-MeV energies, solar neutrinos are the dominant astronomical backgrounds, which are highly localized. An upcoming detector that enables the directional determination of the neutrino flux is thus ideal to observe the diffusive neutrino component from decaying f{\'e}eton DM. Recently, Borexino has successfully made the first direction determination at sub-MeV energies using hybrid light signals (both liquid scintillator and Cherenkov lights)~\cite{BOREXINO:2021efb}. For the high end of f{\'e}eton mass, this directionality technique can already be applied to filter out the solar contamination and can be improved with a new type of liquid scintillator detector~\cite{Guo:2017nnr}. At lower energies, other techniques are under developed such as those for future DM detectors~\cite{Ahlen:2010ub,2010IJMPA..25....1A,2011JPhCS.309a2014S}.   Besides the direction dependence, solar neutrinos exhibit annual modulation which has allowed an independently measure the eccentricity $\epsilon$ of the Earth orbit~\cite{Appel:2022kbr}. Alternatively, a precise determination of $\epsilon$ with an independent method would allow to use this annual modulation as an additional information to remove the flux of solar neutrinos. 


{\bf Discussion and Conclusions\;--} Motivated by the success of the seesaw mechanism and the leptogenesis, we have proposed the $B$-$L$ gauge boson, the {\it f{\'e}eton}, as the dominant DM component in the universe (see e.g.\ Ref.~\cite{Choi:2020dec}). Our f{\'e}eton DM scenario shows various self-consistent features which are worth being emphasized.

First of all, f{\'e}eton is the massive gauge boson naturally equipped in the $B$-$L$ symmetry breaking -- the essential component in the seesaw mechanism and leptogenesis. One important parameter of the theory is the $B$-$L$ symmetry breaking scale $V$. Surprisingly, both the considerations over the neutrino mass scale and the cosmological constraints shown in Fig.~\ref{fig:ParamSpace} (lifetime and WDM constraints) lead to a similar range for the symmetry breaking scale $V\gtrsim10^{12}\,$GeV. At the same time, the gauge coupling constant $\gBL$ is required to be small (e.g., $\gBL\sim10^{-19}$), which also gives a small (sub-MeV) f{\'e}eton mass. This allows f{\'e}eton to decay into active neutrinos but not other SM particles, thus avoiding the strong constraints on decaying DM models from the searches of their by-products. 

Moreover, we have shown that such a consistent f{\'e}eton DM scenario predicts a nontrivial neutrino signal from the f{\'e}eton decay that is potentially detectable by future low-energy neutrino experiments. The detection of such a signal will allow us to verify the existence of the $B$-$L$ extension to the SM. For instance, once the signal from the Galactic neutrino flux is resolved and reconstructed from observations, it would directly provide the value of the f{\'e}eton mass $\mA$ from the position of the peak, while its integrated flux would reveal the value of the gauge coupling constant $\gBL$ independently of $\mA$. Indeed, the $B$-$L$ breaking scale $V=\mA/(2\gBL)$ could in principle be determined solely by measuring the energy and the amplitude of the peak flux. If the inferred value of the breaking scale falls within the range $V=(10^{12}\,\textrm{--}\,10^{16})\,$GeV, it would be a smoking gun for the existence of the seesaw mechanism, leptogenesis, as well as for the f{\'e}eton being indeed the DM particle. Such a test will be strengthened by the detection of the broad-spectrum extragalactic signal.

Besides the above tests to tackle the existence of a $B$-$L$ symmetry breaking, the neutrino flux from f{\'e}eton decay can also be potentially used to probe some long-lasting cosmological and astrophysical problems. For instance, if a combination of both the Galactic and the extragalactic signals is observed, it would offer a novel way to test cosmological models using neutrino astronomy.

In fact, since the distance-redshift relation behaves differently during the late cosmic acceleration era than that during the early matter-dominated era, the spectrum of the extragalactic signal would peak at a lower energy scale than $\mA/2$ due to cosmological redshift, see Fig.~\ref{fig:NuFlux}. Assuming the standard cosmological model, the ratio in energy between the two peaks (extragalactic and Galactic) is
\begin{equation}\label{eq:peak-ratio}
    \frac{E_{\nu,{\rm peak}}^{\rm eg}}{E_{\nu,{\rm peak}}^{\rm \textsc{g}al}}=\left(\frac{1-\Omega_{\Lambda}}{2\Omega_{\Lambda}}\right)^{1/3}\,.
\end{equation}
Reading the relative location of the two energy peaks would then lead to an independent test of the standard cosmological model, independently of the fraction of decaying DM component, the Galactic DM halo profile, or if photons were the dominant component from DM decay.

Also, if the morphology of the Galactic signal can be resolved by a future neutrino detector, it could directly reveal the inner part of the Galactic DM profile. So far, there is an uncertainty about whether the inner region of MW has a cuspy or cored DM profile~\cite{Nesti:2013uwa}. Studies that exploit the dynamics of stars are limited by the fact that the inner Galactic region is dominated by baryons. DM decay signals, on the other hand, directly trace the DM density distribution and would exhibit a pattern that is more concentrated at GC~\cite{Asaka:1998ju,Blasi:2001hr}. The bottom panel of Fig.~\ref{fig:NuFlux} shows distinct features of the neutrino flux within a $\sim30^\circ$ area from GC between the two DM profiles we consider: for a cuspy NFW profile, the neutrino flux rises sharply towards GC, while it flattens out for a cored Burkert profile. 

Note, that the uncertainty in the inner shape of the Galactic DM profile does not qualitatively change our main conclusions, because both a cored Burkert profile and an NFW profile give similar full-sky fluxes and only slightly affect the inference of $\gBL$. The neutrino flux from f{\'e}eton decay can then be used to test the scenario of a new $B$-$L$ symmetry breaking independently of the Galactic DM distribution.

In this paper, we have restricted the presentation to a $B$-$L$ gauge field model. However, the discussions can be generalized to include the hypercharge~\cite{Okada:2020evk}. As long as f{\'e}etons have a non-negligible coupling to neutrinos, our conclusions would not be sensibly modified even in the presence of a kinetic mixing between the f{\'e}eton and the SM photon.

\section{Acknowledgments}

\begin{acknowledgments}
We thank Gongjun Choi for pointing out Ref.~\cite{Graham:2015rva} where the inflationary production of vector DM is discussed. T.\ T.\ Y.\ is supported in part by the China Grant for Talent Scientific Start-Up Project and by Natural Science Foundation of China (NSFC) under grant No.\ 12175134 as well as by World Premier International Research Center Initiative (WPI Initiative), MEXT, Japan. D.\ L.\ X.\ thanks the NSFC grant No.~12175137 on ``Exploring the Extreme Universe with Neutrinos and Multi-messengers'' and the Double First Class start-up fund provided by Shanghai Jiao Tong University.
\end{acknowledgments}

\bibliographystyle{apsrev4-1}
\bibliography{refs}

\end{document}